\documentclass[preprint,proceedings]{rmaa}

\suppressfulladdresses 

\usepackage{paralist}

\usepackage{psfrag,color}
\usepackage{epsf}
\usepackage{psfig}
\usepackage{lscape}

\newcommand{\kms}{\ensuremath{{\rm km\,sec^{-1}}}}                   
\newcommand{\msun}{\ensuremath{\mathit{M}_{\odot}}}                  
\newcommand{\msunyr}{\ensuremath{\mathit{M}_{\odot}{\rm yr}^{-1}}}   
\newcommand{\lsun}{\ensuremath{\mathit{L}_{\odot}}}                  
\newcommand{\zsun}{\ensuremath{\mathit{Z}_{\odot}}}                  

\newcommand{\mdot}{\ensuremath{\dot{M}}}                             
\newcommand{\teff}{\ensuremath{\mathit{T}_{\rm eff}}}                
\newcommand{\vinf}{\ensuremath{v_{\infty}}}                          

\title{Mass loss from Luminous Blue Variables and quasi-periodic modulations of radio supernovae}

\author{
  Jorick S. Vink,\altaffilmark{1} 
  Rubina Kotak\altaffilmark{2}
  }

\altaffiltext{1}{Keele University, Astrophysics Group, Lennard-Jones Labs, ST5 5BG, Staffordshire, United Kingdom}
\altaffiltext{2}{European Southern Observatory, Karl-Schwarzschild Str. 2, Garching bei M\"unchen, D-85748, Germany}    

\shortauthor{Vink \& Kotak}
\shorttitle{Mass loss from Luminous Blue Variables and Radio SNe}

\listofauthors{Vink, Jorick S., \& Kotak, Rubina}
\indexauthor{Vink, Jorick S.}
\indexauthor{Kotak, Rubina}

\abstract{Massive stars, supernovae (SNe), and long-duration gamma-ray bursts (GRBs) 
have a huge impact on their environment. Despite their importance, a comprehensive 
knowledge of which massive stars produce which SN/GRB is hitherto lacking. 
We present a brief overview about our knowledge of mass loss in the 
Hertzsprung-Russell Diagram (HRD) covering evolutionary phases of 
the OB main sequence, the unstable Luminous Blue Variable (LBV) stage, and 
the Wolf-Rayet (WR) phase. 
Despite the fact that metals produced by ``self-enrichment'' in WR atmospheres 
exceed the initial -- host galaxy -- metallicity, by orders of magnitude, a particularly 
strong dependence of the mass-loss rate on the {\it initial} metallicity is found 
for WR stars at sub-solar metallicities (1/10 -- 1/100 solar). This provides a significant 
boost to the collapsar model for GRBs, as it may present a viable mechanism to 
prevent the loss of angular momentum by stellar winds at low metallicity, whilst strong Galactic 
WR winds may inhibit GRBs occurring at solar metallicities. 
Furthermore, we discuss recently reported 
quasi-sinusoidal modulations in the radio lightcurves of SNe 2001ig and 2003bg.
We show that both the sinusoidal behaviour and the recurrence 
timescale of these modulations are consistent with the predicted mass-loss behaviour 
of LBVs, and we suggest LBVs may be the progenitors of some core-collapse SNe.}

\addkeyword{Stars: Mass loss}
\addkeyword{Stars: Stellar winds}
\addkeyword{Stars: Wolf-Rayet}
\addkeyword{Stars: Luminous Blue Variable}
\addkeyword{Supernovae}

\begin{document}

\maketitle

\section{Introduction}

Massive star winds and core-collapse supernovae (SNe) have a huge influence 
on their environments by driving the chemical evolution of galaxies and 
shaping the interstellar medium over all cosmological epochs, since 
the very first stars came into existence. 
Despite their importance, the lives and deaths of massive 
stars are poorly understood. Despite theoretical progress \citep[e.g.][]{hirschi04}, 
it is not known with any
degree of certainty which massive stars produce which SNe/GRB. 

While progress is being made in the direct identification of 
SN progenitors by searching for these in pre-explosion images 
\citep[e.g.][]{smartt02,vdyk03}, current progenitor masses appear
to be limited to stellar masses not significantly greater than
$\sim$10-15$\msun$, likely as a result of the initial mass function.

The evolution of more massive stars ($M$ $>$ 40 \msun) is 
largely unconstrained, but it is generally accepted that mass loss drives 
these objects through the O star, Luminous Blue Variable (LBV), and Wolf-Rayet (WR)
phases \citep[e.g.][]{chiosi86}.
Mass loss also determines the stellar mass before collapse, and is therefore relevant 
for the type of compact remnant that is left behind (i.e. neutron star or 
black hole). This process is expected to depend on the metal content ($Z$) of the 
host galaxy \citep[e.g.][]{eldridge06}. As WR stars are the likely progenitors 
of long-duration GRBs \citep{woos93}, the strength of WR winds as a function of 
$Z$ is especially relevant for setting the threshold $Z$ for forming GRBs.

Furthermore, massive stars explode in environments that have been modified by 
mass loss from the progenitor. The SN ejecta interact first
with this circumstellar material before interacting with
interstellar material. We might therefore expect that the different 
wind properties over the lifespan of a massive star be imprinted onto the 
resulting circumstellar media (CSM), and we would expect 
these differences to be seen in the interaction between the SN ejecta 
and surrounding material. By quantifying these differences one may be able to
constrain the evolutionary phase of the exploding object. 

Over the last decades, radio observations of SNe have provided a means with 
which to constrain the density of the CSM around core-collapse SNe. 
The inferred mass-loss rates from modelling of most radio SN light curves yield values of 
$\mdot$ $\sim$$10^{-6}$ -- $10^{-4}$$\msunyr$ 
\citep[e.g. compilation in][]{weiler02}. Unfortunately, these average mass-loss rates
are generally only accurate to within a factor of $\sim$10, and are typical of almost all 
types of massive star, making it difficult to pin down the evolutionary phase during
which core-collapse occurred.

However, a small subset of radio SNe show quasi-periodic modulations in their radio 
lightcurves. We argue that this type of modulation may be the result 
of an LBV that underwent S~Doradus variations which entailed opacity changes in the 
wind-driving region, resulting in varying mass-loss rates \citep{vink02,kotak06}.

Given the crucial role that mass loss plays for massive star evolution,
we briefly discuss recent mass-loss predictions in order of decreasing temperature: 
WR stars $\rightarrow$ OB supergiants $\rightarrow$ LBVs (Sects.~2~-~4). 
In Sects.~5 and 6, we link our knowledge of mass loss to 
certain types of radio SNe, and discuss whether LBVs may explode 
in Sect.~7.

\section{Wolf-Rayet mass-loss rates as a function of metal content}
\label{s_WR}

In recent years, it has become clear that long-duration gamma-ray bursts (GRBs) are 
associated with the explosion of a massive star, providing 
impetus to the collapsar model \citep{macf99}. The model works best if the 
progenitor fulfils the following two criteria: (i) the absence of a thick 
hydrogen envelope (enabling the jet to emerge), and (ii) rapid rotation of 
the core (allowing a disk to form). This may point 
towards a rapidly rotating WR star. 

\begin{figure}[!t]
\includegraphics[width=\columnwidth]{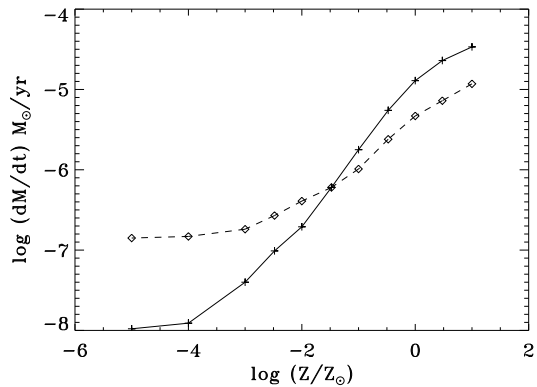}
\caption{Mass loss versus initial $Z$ for late-type nitrogen-rich (WN) stars (solid line) and 
carbon-rich WC stars (dashed line). 
Note that metal self-enrichment is accounted for, but does 
not enter in our expression of $Z$. See \citet{vink05} for details.}
\label{f_wnwc}
\end{figure}

WR stars are believed to be the result of mass-loss during earlier 
evolutionary phases (the ``Conti'' scenario \citep{conti76}), while in a 
complementary scenario, the removal of the thick hydrogen envelope 
may be due to a companion. Recently, an alternative scenario for producing 
a GRB progenitor has gained popularity \citep{yoon05,woos06}: when a star rotates rapidly,
it may mix ``quasi homogeneously'', and the object may not develop the 
classical core-envelope structure, but remain small. A potential problem for producing 
a GRB in this scenario is that Galactic WR stars have strong stellar winds which may remove 
the angular momentum \citep{langer98}, making it challenging to produce a GRB.

\begin{figure}[!t]
  \includegraphics[width=\columnwidth]{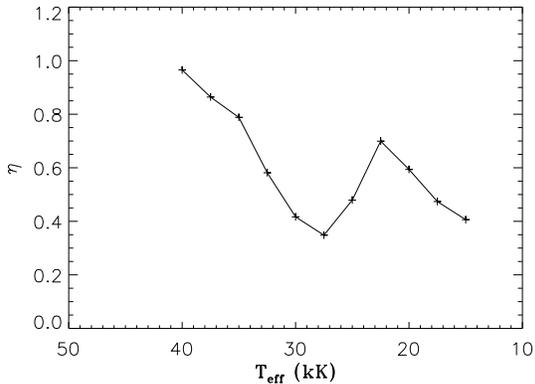}
 \caption{Wind efficiency $\eta = (\dot{M}\vinf)/({L_*/c})$ 
as a function of effective temperature. The predictions are taken from \citet{vink00}.
Note the presence of the bi-stability jump around 25 kK, where $\eta$ increases 
as Fe recombines to Fe {\sc iii}. }
\label{f_eta}
\end{figure}

This problem might be overcome if WR winds are weaker at low $Z$, so 
the question is: ``are the winds of WR stars $Z$-dependent?'' and if 
so, ``how strong is this dependence?'' 
The dense winds of WR stars are likely driven by radiation pressure 
\citep{nugis02, graf05}, just like their less extreme O star counterparts. 
 
This need not imply that WR winds depend on metal content, as WR stars produce 
copious amounts of metals such as carbon (in WC stars).
If, on the one hand, these self-enriched elements dominate the driving (by their sheer 
number of particles), one would expect WR winds to be independent of their initial $Z$ 
and the requirements of the collapsar model may never be met. If, on the 
other hand, iron (Fe) is predominantly 
responsible for the driving \citep[as in O stars;][]{vink01}, WR winds might indeed be 
less efficient in low $Z$ galaxies. 

To address this issue regarding the $Z$-dependence of WR winds, \citep{vink05} 
computed mass-loss rates for late-type WN and WC stars as a function of the initial metal 
content (representative of the host galaxy $Z$). 
The results are shown in Fig.~\ref{f_wnwc}. 
For a discussion of the flattening in the mass-loss-$Z$ dependence for initial metallicities 
below log ($Z/\zsun$) $= -2$ and potential consequences for the first stars (Pop {\sc iii}), the 
reader is referred to \citet{vink06}, but for the $Z$ range 
down to log ($Z/\zsun$) $= -2$, the mass loss is found to drop steeply, as \mdot\ $\propto$ $Z^{0.85}$,  
for the WN phase - where WR stars spend most of their time. This inefficiency of WR mass loss
at subsolar $Z$ may prevent the loss of stellar angular momentum, and may provide 
a boost to the collapsar model.

\section{Mass loss from OB stars: absolute rates and the bi-stability jump}
\label{s_ob}

We now switch from a discussion of $Z$-dependent mass loss to one of  
\teff-dependent mass loss. We describe the expected wind properties 
in terms of their wind efficiency number $\eta = (\dot{M}\vinf)/({L_*/c})$, a measure 
for the momentum transfer from the photons to the ions in the
wind. \citet{vink00} computed wind models
as a function of effective temperature (Fig.~\ref{f_eta}).
The overall behaviour is one of decreasing $\eta$ with decreasing \teff\ due to a 
growing mismatch between the wavelengths of the maximum opacity (in the UV) and 
the flux (gradually moving towards longer wavelengths). 
The behaviour changes at the ``bi-stability jump'' (BSJ; e.g. Lamers et al. 1995), 
where $\eta$ {\it increases} by a factor of 2-3, as Fe {\sc iv}
recombines to Fe {\sc iii} \citep{vink99}. 

Recent mass-loss studies \citep{trundle05, crow06} have reconfirmed discrepancies 
between empirical mass-loss rates and predictions for B supergiants \citep{vink00}. 
Discrepancies have also been reported for O stars \citep{bouret03,ful06}, and it is as yet 
unclear whether the reported discrepancies for B supergiants are due to model 
assumptions (e.g. the neglect of wind clumping) or the physical reality of the BSJ. 
The most accurate way to derive \mdot\ is believed to be through radio observations.
Intriguingly, \citet{benag07} present empirical radio mass-loss rates as a function
of effective temperature that resemble the mass-loss efficiency behaviour  
predicted by \citet{vink00}. This may well be the first evidence of the presence of 
a mass-loss BSJ at the boundary between O and B supergiants. 
The relevance for stellar evolution is that when massive stars evolve at constant luminosity
towards lower \teff, they are anticipated to cross 
the BSJ. Interestingly, LBVs brighter than log ($L/\lsun$) $= 5.8$ 
(see Fig.~\ref{f_smith}). 
are expected to encounter it continuously - on timescales of their photometric 
S~Doradus variability, discussed in the next section.

\section{Mass loss from Luminous Blue Variables}
\label{s_lbv}

\begin{figure}[!t]
\includegraphics[width=\columnwidth]{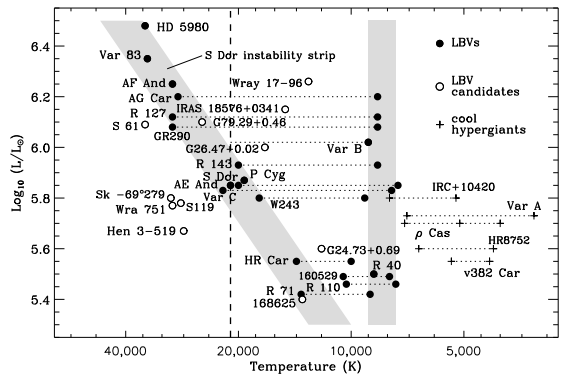}
\caption{The LBVs in the HRD. The shaded areas represent the S~Doradus instability strip (diagonal) and the position
of the LBVs during outburst (vertical). The dashed vertical line at 21~000 K indicates the position of the 
bi-stability jump. The figure is taken from \citet{smith04}.}
\label{f_smith}
\end{figure}

\begin{figure}[!t]
\includegraphics[width=\columnwidth]{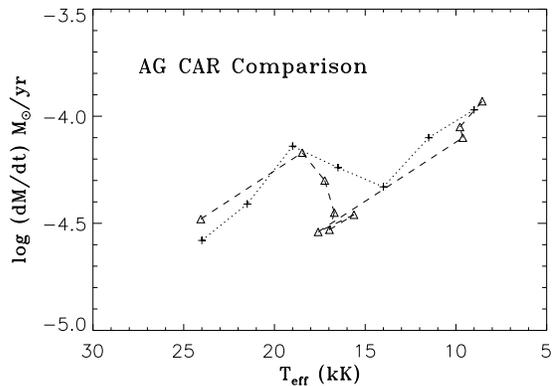}
\caption{Predicted (dotted line) and empirical (dashed line) mass-loss rates versus \teff\ 
for the LBV AG~Car. Note that both the qualitative behaviour and the amplitude of the 
mass-loss variations are well reproduced. See \citet{vink02} for details.}
\label{f_ag}
\end{figure}

LBVs are unstable massive stars in the upper part of the HRD \citep[e.g.][]{hd94}. 
As can be seen in Fig.~\ref{f_smith}, the ``classical'' LBVs, like AG Car, are anticipated 
to cross the BSJ at $\sim$ 21 000 K. One of the defining characteristics for LBVs is their 
S~Doradus (SD) variation of $\sim$1 -- 2 mag on timescales of years 
(short SD phases) to decades (long SD phases)
\citep{vgenderen01}. \citet{vink02} computed LBV mass-loss rates as a function of \teff - shown 
in Fig.~\ref{f_ag}. Overplotted are the empirical H$\alpha$ mass-loss rates for AG Car 
\citep{stahl01}, which vary on the timescales of the photometric S~Doradus variability.
Although the agreement is not perfect (see \citet{vink02} for a discussion), the amplitude of 
the predicted variability fits the observations well, and most importantly the overall behaviour 
appears to be very similar, and may indeed be explained in terms of the physics of the BSJ. 
This bi-stable behaviour in an individual stellar wind \citep{pp90} causes 
the star to flip back and forth between two states: that of a low mass 
loss, high-velocity wind, to a high mass-loss, low velocity wind.
The wind density ($\propto \mdot/\vinf$) 
would therefore be expected to change by a factor 
of $\sim$2~$\times$~$\sim$2, i.e. $\sim$4 on the timescale of the SD variations. In the absence of any other material around 
the star, this would result in a pattern of concentric shells of varying density.

\section{Radio supernovae and progenitor mass loss}
\label{s_radio}

Radio SNe (RSNe) lightcurves and the model for SN interaction with 
the surrounding circumstellar material has been reviewed by \citet{weiler86}.
The radio emission is due to 
non-thermal electrons, while the absorption may be due to both synchrotron 
self absorption as well as free-free absorption \citep{chevalier82,fb98}. 
Examples of the rise, peak, and power-law decline of radio lightcurves are 
shown in Fig.~\ref{f_soder}. 
(The episodic bumps at late time are discussed in Sect.~\ref{s_quasi})

The model constrains the wind density and thus the ratio of \mdot\ to the terminal 
wind velocity ($\vinf$): $\rho \propto \mdot/\vinf r^2$. Assuming $\vinf$, 
\citet{weiler02} list \mdot\ values in the range $10^{-6}$--$10^{-4}$ \msunyr. Fortunately, 
these values agree with mass-loss predictions, but are broadly representative for massive 
stars over almost all post-main sequence evolutionary phases, making it hard to 
infer the progenitor from radio lightcurves alone, unless these lightcurves betray
their progenitor in some another way.


\section{Quasi-periodic oscillations in radio SNe lightcurves}
\label{s_quasi}

\begin{figure}[!t]
\includegraphics[width=\columnwidth]{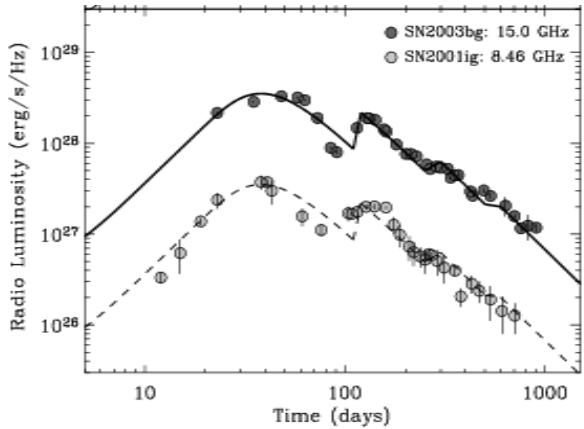}
\caption{Radio luminosity versus time for two strikingly similar recent SNe: 2001ig and 2003bg. 
Note the quasi-sinusoidal modulations during the power-law decline phase. Taken from 
\citet{soderberg05}.}
\label{f_soder}
\end{figure}

A number of recent RSNe have shown sinusoidal modulations in their radio lightcurves, in particular
SN~2001ig \citep{ryder04} and SN~2003bg \citep{soderberg05} are strikingly 
similar in terms of both amplitude and variability timescale (see Fig.~\ref{f_soder}). 
The recurrence timescale $t$ of the bumps is $\sim$ 150 days. 
Using Eq.~(13) from \citep{weiler86}:

\begin{equation}
\Delta P~=~\frac{R_{\rm shell}}{v_{\rm wind}}~=~\frac{v_{\rm ejecta}~t_{\rm i}}{v_{\rm wind}~m} \left(\frac{t}{t_{\rm i}}\right)^m
\label{eq:period}
\end{equation}
where $m$ is the deceleration parameter (here $m$ = 0.85) and $t_{\rm i}$ is the time of measurement of the ejecta 
velocity relative to the moment of the explosion.
Assuming $v_{\rm wind}$ = 10--20\,\kms, typical wind velocities for 
red (super)giants, \citep{ryder04} found a period $P$ between successive 
mass-loss phases that was too long for red (super)giant pulsations (100s of days, see however \citet{heger97}), 
but too short for thermal pulses (10$^2$--10$^3$ years). They therefore 
invoked an edge-on, eccentric binary scenario involving a WR-star and a massive companion. 
One of the main differences between LBV and red giant winds is that 
LBV winds are about 10 times faster. If the progenitor of 
SN~2001ig were an LBV, the expected period between successive 
mass-loss episodes would be $\Delta P \sim$\,25\,yr 
(for an assumed $v_{\rm wind}$~=~200\,\kms), consistent with the long SD phase \citep{kotak06}.

\citet{soderberg05} infer density enhancements 
of a factor of $\sim 2$ during the deviations from pure power-law evolution.
They consider a range of options that might account for the modulations, but they 
favour a single-star progenitor model of a WR star that underwent episodes of 
intensified mass loss. However, they do not specify the physical 
mechanism that gives rise to these periods of enhanced mass loss. Our 
SD mechanism for LBVs may alleviate this shortcoming.

\section{Discussion: do LBVs explode?}
\label{s_concl}

Are LBVs viable SNe progenitors? 
It may be relevant that both SNe ~2001ig and 2003bg are ``transitional'' objects.
SN~2001ig was initially classified as type II (showing H lines) but metamorphosed 
into a type Ib/c object (no H lines, weak He lines) about 9 months later.
This suggests that it has lost most of its H-rich envelope. SN~2003bg however was first 
classified as a type Ic, but within a month the spectrum evolved into a type II SN. This 
transitional behaviour hints at the fact that their progenitors are intermediate evolutionary 
objects: H-rich compared to OB/red (super)giants, but H-poor compared to WR stars.
LBVs are likely candidates. 

Recently there has been much discussion regarding clumping in the winds of O stars.
The value for the clumping factor is very much an open issue. 
\citet{mokiem07} show that if the empirical H$\alpha$ rates are overestimated
by a factor of two due to clumping, these empirical rates are in good agreement with
the mass-loss predictions of \citet{vink00,vink01}, and consequently our current knowledge of 
massive star evolution is not anticipated to be affected by clumped winds. 
If however the wind clumping factor would be {\it significantly} larger than a factor two/three 
(as has been suggested by UV analyses), this could have severe
implications for massive star evolution. 
One consequence might be that giant LBV eruptions ($\eta$ Car type eruptions, not
the typifying SD variations) dominate the integrated 
mass loss during evolution \citep{smith06}.
An alternative scenario could be that post-main sequence stars do not become WR stars, 
but explode early -- during their LBV phase.

Here, we have presented indications that at least those SNe that show 
quasi-periodic modulations in their radio lightcurves 
might have LBV progenitors \citep{kotak06}. 
It has also been speculated that LBVs may be the generic progenitors 
of type IIn SNe \citep{gal06}, however it may be more relevant to discuss
IIn SNe as a ``phenomenon'' describing SN ejecta expanding into a dense CSM than 
a one-to-one correlation to a particular progenitor \citep{kotak04}. 
Nevertheless, some fraction of type IIn SNe may well have
LBV progenitors although the observational evidence remains elusive.

It is relevant to note that the LBV candidate HD168625 is embedded in a 
bipolar-shaped nebula that resembles the triple-ring system around SN1987A.
This similarity could hint that the progenitor of 1987A (i.e. 
the blue supergiant Sk-69 202) underwent an LBV giant eruption before it exploded
\citep{smith07}.

Future mass-loss predictions are anticipated to play an important role 
in obtaining knowledge about the lives and deaths of massive stars.


\end{document}